\documentclass[adp,fleqn]{w-art}
\usepackage{times}
\usepackage{w-thm}
\usepackage{amssymb}

\usepackage[]{graphicx}
\chardef\bslash=`\\ 

\begin{document}
\DOIsuffix{theDOIsuffix}
\Volume{XX} \Issue{X} \Copyrightissue{XX} \Month{XX}
\Year{20XX} 
\pagespan{1}{}
\Receiveddate{1 September 2009} 
\Accepteddate{5 September 2009}



\title[Charge transport in nonpolar organic materials]{Quadrupolar glass as a model for charge carrier transport
in nonpolar organic materials}


\author[S.V. Novikov]{Sergey V. Novikov\footnote{E-mail: {\sf novikov@elchem.ac.ru}, Phone: +7\,495\,952\,2428,
     Fax: +7\,495\,952\,5308}}
\address[]{A.N. Frumkin Institute of Physical Chemistry and
Electrochemistry, Leninsky prosp. 31, 119991 Moscow, Russia}
\begin{abstract}
Monte Carlo simulation of the charge carrier transport in
disordered nonpolar organic materials has been carried out. As
a suitable model we considered the model of quadrupolar glass.
A general formula for the temperature and field dependence of
the mobility
 was suggested. A comparison with experimental data
 has been carried out.
\end{abstract}
\maketitle





\section{Introduction}
\label{sect1}

Characteristic feature of disordered organic materials is a
long range spatial correlation in the distribution of random
energies $U(\vec{r})$ of transport sites
\cite{Novikov:14573,Dunlap:542}. This particular feature stems
from two fundamental properties of typical organic glasses:
first, they have very low concentration of intrinsic free
carriers, and, second, organic glasses usually have high
concentration of dipoles and quad\-rupoles. Due to almost zero
concentration of free carriers and lack of screening such
molecules provide long range contribution to the overall
energetic disorder for charge carriers. Long range sources
inevitably lead to the strong spatial correlation of the
resulting random energy landscape. Experimental data and
realistic models of organic glasses suggest that the density of
states usually has a Gaussian form
 with typical rms disorder $\sigma\simeq
0.1$ eV \cite{Bassler:15,Dieckmann:8136,Novikov:877e}. Hence,
the only relevant characteristic of the random energy
$U(\vec{r})$ is the binary correlation function $C(\vec{r}) =
\left<U(\vec{r})U(0)\right>$. Exact solution of the 1D
transport model shows that this function directly dictates a
functional form of the drift mobility field dependence $\mu(E)$
\cite{Dunlap:542}. For this reason organic glasses with
different $C(\vec{r})$ must have different mobility field
dependences.

For example, if correlation function decays as a power law
\begin{equation}
C(\vec{r})=A\sigma^2\left(\frac{a}{r}\right)^n
\label{Cn}
\end{equation}
(here $a$ is a minimal distance between transport sites), then
the 1D transport model gives in the most important case of
strong disorder $\sigma\beta\gg 1$ \cite{Novikov:2532}
\begin{equation}
\ln\mu/\mu_0\approx-2\alpha
a-\sigma^2\beta^2+\left(1+\frac{1}{n}\right)
\sigma\beta\left(An\sigma\beta\right)^{\frac{1}{n+1}}\left(\frac{eaE}{\sigma}\right)^{\frac{n}{n+1}},
\hskip10pt \beta=1/kT,
\label{1D}
\end{equation} 	
where $\alpha$ is a radius of the decay of the wave function of
the transport level. This particular case is of the utmost
importance, because in the simplest model of polar disordered
organic materials (the model of dipolar glass
\cite{Novikov:14573}) the correlation function decays as $1/r$,
and in the simplest model of nonpolar disordered organic
materials (the model of quadrupolar glass (QG)
\cite{Novikov:89}) $C(\vec{r})\propto 1/r^3$. Simplest
realizations of the models could be considered as regular
lattices with sites occupied by randomly oriented dipoles or
quadrupoles, respectively.

According to the result of the 1D model, in polar organic
materials the mobility field dependence should obey the law
$\ln\mu\propto E^{1/2}$, and in nonpolar materials the
dependence is $\ln\mu\propto E^{3/4}$. This result shows that
the earlier belief, promoted by the Gaussian Disorder Model
(GDM) \cite{Bassler:15}, that there is a universal mobility
field dependence in disordered organic materials is, in fact,
wrong. The GDM assumes that there is no spatial correlation in
the random energy landscape and corresponding mobility field
and temperature dependence could be formally considered as the
particular limiting case of (\ref{1D}) for $n\rightarrow
\infty$. 3D Monte Carlo simulation generally supports the
result of the simplified 1D model and typically only modifies
the numerical coefficients in (\ref{1D}) \cite{Novikov:4472}.

At the same time, experiment indicates that the universal
Poole-Frenkel (PF) mobility field dependence $\ln\mu\propto
E^{1/2}$ is usually a good approximation for the description of
the time-of-flight (TOF) data.  The major goal of this paper is
to present results of the extensive 3D Monte Carlo simulation
for the QG model and discuss how well they fit the TOF data for
charge transport in nonpolar organic materials, and how the
intrinsic mobility field dependence $\ln\mu\propto E^{3/4}$ can
be reconciled with the seemingly universal Poole-Frenkel
dependence $\ln\mu\propto E^{1/2}$, found in experiments.

\section{Quadrupolar glass model: transport simulations and results}
\label{sect3}

Simulations have been done for a lattice QG model with the size
of basic sample of $128\times 128\times 128$ sites of a simple
cubic lattice with the lattice scale $a$ and periodic boundary
conditions imposed. Particular distributions of $U(\vec{r})$
have been generated in the usual way. There  is no correlation
among the fluctuations in momentum space $U(\vec{k})$ for
different $\vec{k}$, so we generated random Gaussian field
$U(\vec{k})$ for the quadrupolar correlated function, which
with very good accuracy can be described by (\ref{Cn}) with
$n=3$ and $A\approx 0.5$ for $r\ge a$, and, of course,
$C(0)=\sigma^2$ \cite{Novikov:89}, and then calculated Fourier
transform to get $U(\vec{r})$. Carrier hopping has been
simulated using the Miller-Abrahams hopping rate
\begin{equation}
\Gamma_{i\rightarrow j}=\Gamma_0\exp\left(-2\alpha|\vec{r}_j-\vec{r}_i|\right)
\begin{cases}
\exp\left(-\frac{U_j-U_i}{kT}\right),& U_j > U_i,\\
1,&
U_j < U_i,
\end{cases}
\label{MA}
\end{equation}
which is believed to be a good approximation for the hopping
process in organic glasses \cite{Bassler:15}; we used $\alpha
a= 5$, as in \cite{Bassler:15,Novikov:4472}. In (\ref{MA})
random energies $U_i$ include the shift
 from the applied field $E$.

\begin{figure}[htb]
\includegraphics[width=.45\textwidth]{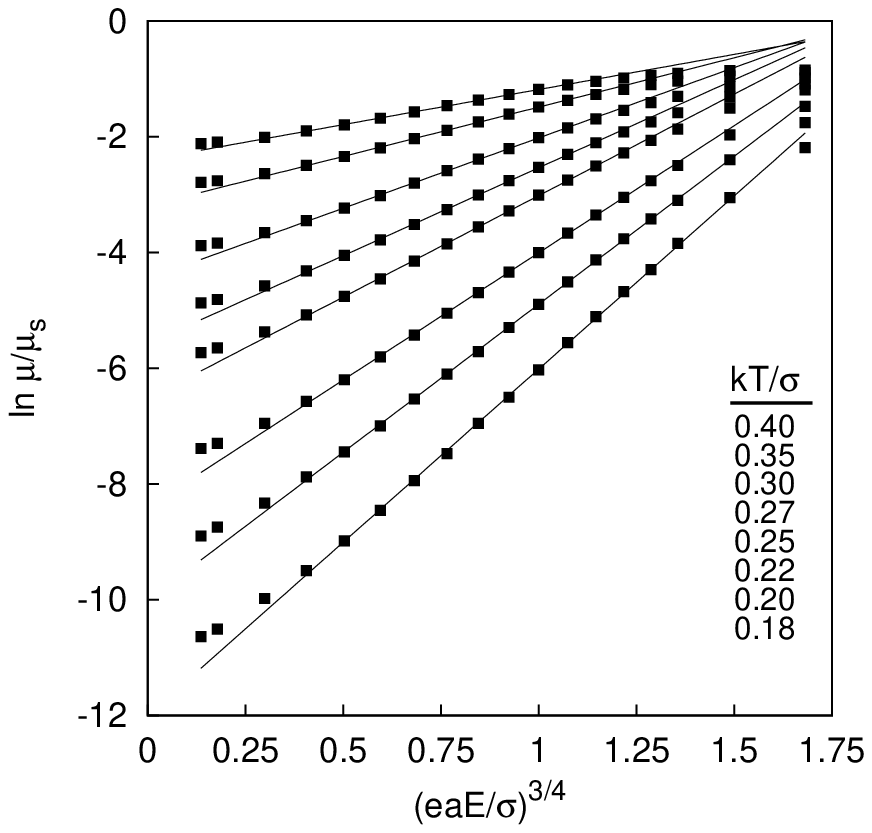}~a)
\hfil
\includegraphics[width=.45\textwidth]{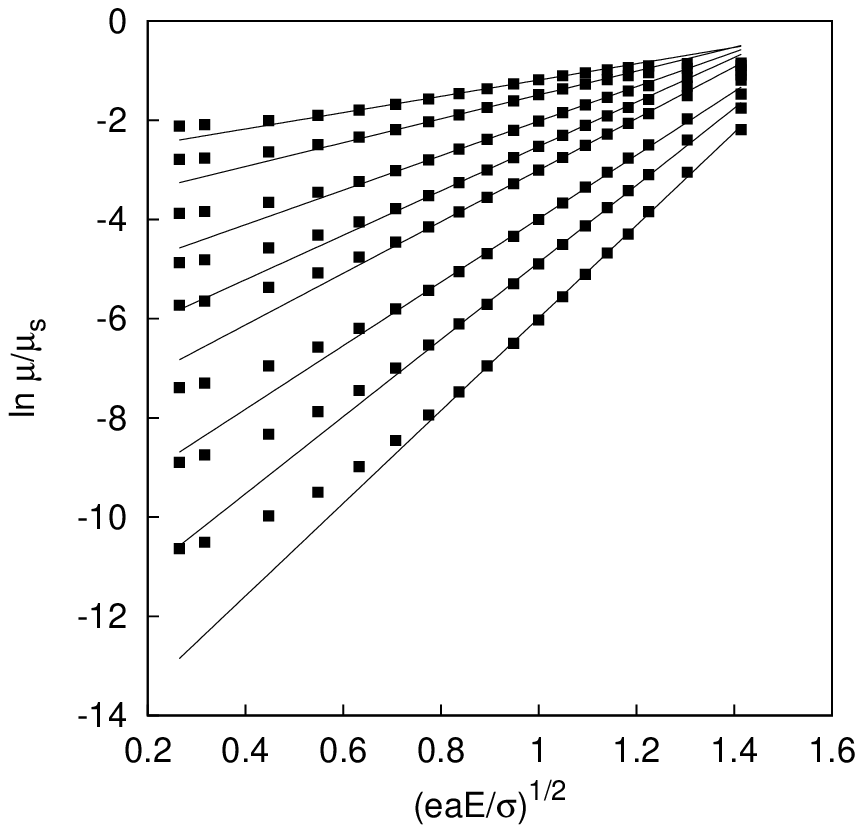}~b)
\caption{Mobility field dependence in the QG model for different
values of $kT/\sigma$ (from the top curve downward);
straight lines in (a)
indicate the fit for (\ref{QGmu}); $\mu_s=\mu_0\exp\left(-2\alpha
a\right)$. Plot of the simulation data
in the usual PF presentation $\ln\mu$ vs $E^{1/2}$ (b) demonstrates
much stronger deviation from the linearity in the weak field
region (straight lines serve as a guide for an eye). If $a\approx 1$ nm and $\sigma\approx 0.1$ eV,
then $eaE/\sigma\approx 1$ for $E=1\times 10^6$ V/cm.}
\label{fig4}
\end{figure}

Details of the Monte Carlo simulation and analysis of the
transport data are almost identical to those described in
\cite{Novikov:4472}. General features of the current transients
are usual: if $\sigma\beta$ is not very large and the transport
layer not too thin, then a well defined plateau is developed,
demonstrating essentially quasi-equilibrium charge transport
with average carrier velocity independent of time. In the
quasi-equilibrium regime the dependence of $\mu$ on $T$ and $E$
is shown in Fig. \ref{fig4}. It was found that the general
field dependence of the quasi-equilibrium mobility closely
follows prediction of 1D model (i.e., $\ln\mu\propto E^{3/4}$),
though the numeric coefficients are different. A
phenomenological relation of $\mu$ on $T$ and $E$ could be
described as
\begin{equation}
\mu =\mu_0\exp\left(-2\alpha
a\right)\exp\left[-0.37\hat{\sigma}^2+C_Q\left(\hat{\sigma}^{5/4}-
\Gamma_Q\right)\left(eaE/\sigma\right)^{3/4}\right],\hskip10pt \hat{\sigma}=\sigma/kT,
\label{QGmu}
\end{equation}
with $C_Q\approx 0.87$ and $\Gamma_Q\approx 1.91$.

\section{Quadrupolar glass: comparison with experimental data}

\begin{figure}[htb]
\includegraphics[width=.45\textwidth]{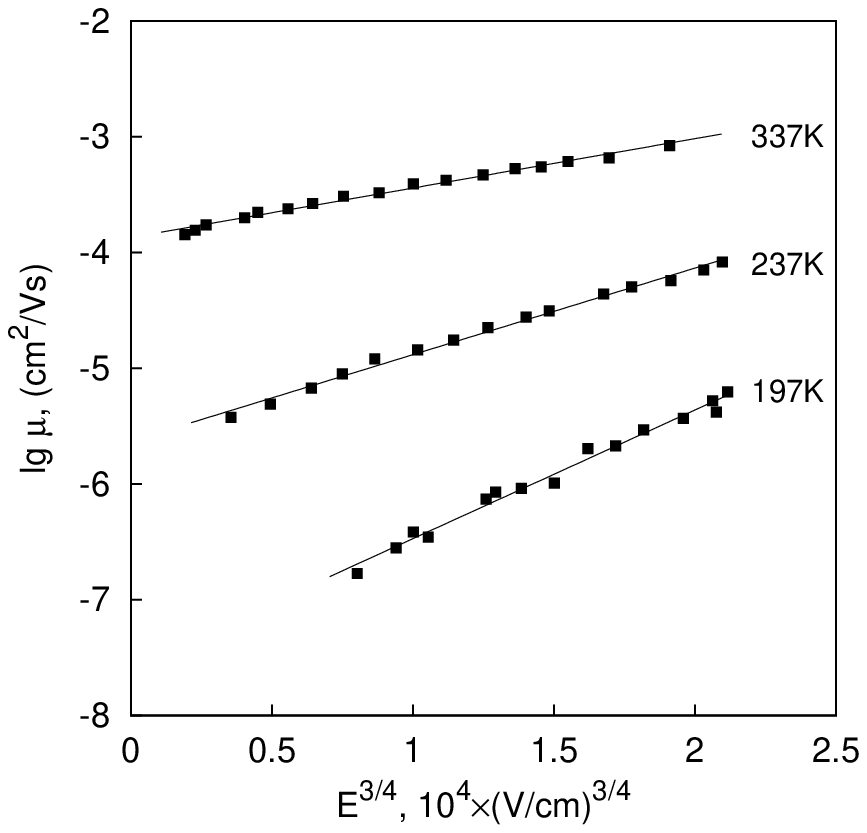}~a)
\hfil
\includegraphics[width=.45\textwidth]{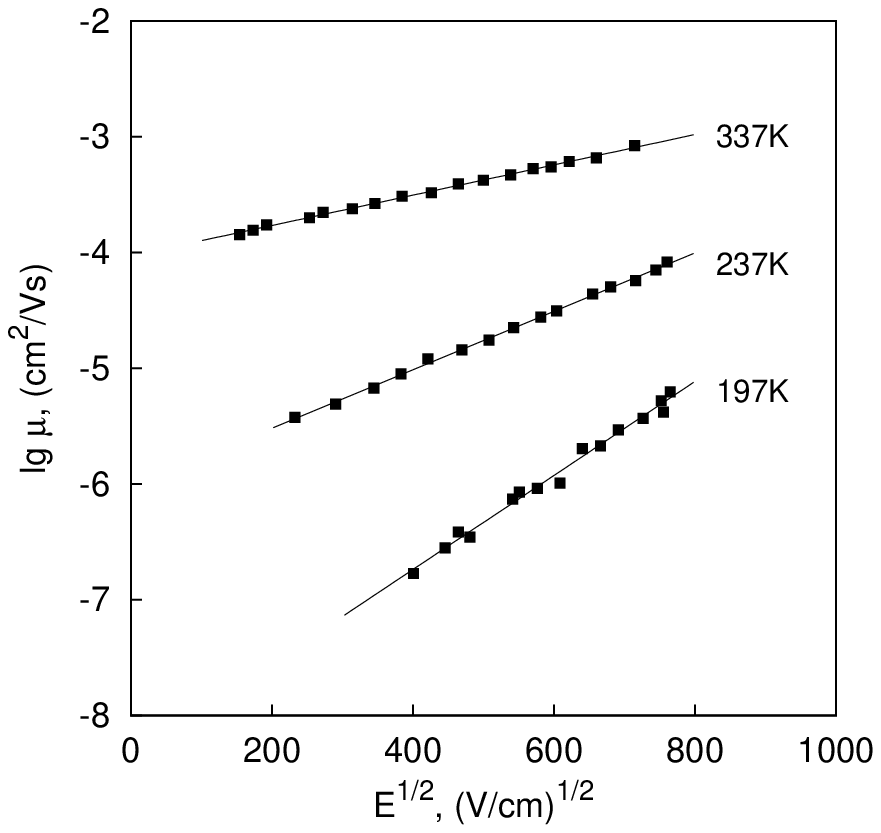}~b)
\caption{Mobility field dependence in polysterene doped with 50\% of
 weakly polar transport dopant TASB
 (bis(di\-tolyl\-amino\-styryl)benzene, dipole moment 0.54 D),
 plotted as $\ln\mu$ vs $E^{3/4}$ in accordance with
  (\ref{QGmu}) and as a usual $\ln\mu$ vs $E^{1/2}$ PF plot.
 Straight lines show the best linear fits and numbers indicate
 the temperature. Both plots demonstrate approximately the same
 linearity. The experimental data has been taken from Figure 3 in
  \cite{Borsenberger:233}.}
\label{fig5}
\end{figure}

Close inspection of Fig.~\ref{fig5}b explains, why the
description of the mobility field dependence in nonpolar
materials in the usual PF presentation $\ln\mu$ vs. $E^{1/2}$
provides a good linear fit in many cases. Indeed, if field
range is not too wide (one order of magnitude or even more
narrow, which is typical for such materials
\cite{tapc,ena,ttb,Borsenberger:233}), then it is nearly
impossible to distinguish the true quadrupolar dependence
(\ref{QGmu}) from the PF one (see Fig. \ref{fig5}).

In this situation the use of the proper method of the mobility
calculation from the TOF curve is extremely important. There
are two most frequently used procedures for the calculation of
$\mu=L/Et$ from the TOF data in double linear current versus
time plot: 1) calculation that uses time $t_i$ determined by
the intersection of asymptotes to the plateau and trailing edge
of the transient and 2) calculation that uses time $t_{1/2}$
for current to reach half of its plateau value; here $L$ is a
thickness of the transport layer. Unfortunately, the first
method is a method of choice for most papers. One can find in
literature the statement that the difference between two
procedures is not very important for determination of the
temperature and field dependence of the mobility. This is not
true. Sometimes the first procedure distorts the functional
type of the mobility field dependence
\cite{Veres:377,Novikov:444}. Procedure that uses $t_{1/2}$ is
much more reliable and generally agrees well with the standard
definition of the mobility as $\mu=v/E$, where $v$ is an
average carrier velocity \cite{Novikov:444}. This very
procedure must be used for the reliable estimation of the
mobility field dependence.

Let us discuss how good is the model of quadrupolar glass for
the description of the transport properties of nonpolar organic
materials. First of all, most transport molecules, polymer
binders or inert dopants do possess dipole moments, albeit
sometimes quite small ones. For example, in a typical nonpolar
polymer matrix, polysterene, a unit of the polymer chain has a
small dipole moment 0.4 D \cite{PS}. Hence, a more proper model
for many nonpolar glasses should be a mixture of the dipolar
and quadrupolar glasses. Unfortunately, such model should be
hardly tractable because the mobility field dependence is a
mixture of dipolar and quadrupolar contributions and individual
contributions cannot be easily separated even in the 1D case
\cite{mix}.

Second, there is another, more fundamental limitation for the
direct applicability of the QG model. In glasses there is no
translational order and, as a consequence, the symmetry is
broken. For this reason molecules could possess induced dipole
moments; for example, in symmetric PPVs, having no permanent
dipole moment, induced dipole moments could reach $\simeq 0.5$
D \cite{yaron}. Induced dipolar disorder is usually smaller
than the quadrupolar disorder, and yet for the weak field
region its contribution can be quite comparable to the
contribution of the dominant quadrupolar disorder
\cite{Novikov:2532}. In this situation no simple relation for
the mobility field dependence can be derived. Very probably,
the only tool suitable to provide the mobility field and
temperature dependence in this complicated situation is a
direct simulation of the materials structure with subsequent
calculation of all site energies and transfer integrals
\cite{nelson}.

\section{Conclusion}
We have shown that the field dependence of the mobility in
nonpolar organic materials differs from the corresponding
dependence in polar organic materials (Poole-Frenkel
dependence). Yet the difference is not very large and this is
the reason why the PF dependence usually fits well experimental
data for nonpolar materials too. Nonetheless, evaluation of the
disorder parameters (i.e., $\sigma$) for nonpolar materials
from the PF mobility dependence can give inaccurate results.

\begin{acknowledgement}
This work was supported by the ISTC grant 3718 and RFBR grant
 08-03-00125.
 \end{acknowledgement}

\end{document}